\documentclass[ reprint]{revtex4-2}

\usepackage{amssymb}   
\usepackage{graphicx}
\usepackage{color,epsfig}
\usepackage{amsmath}
\usepackage{graphicx}
\usepackage{dcolumn}
\usepackage{bm}
\usepackage[T1]{fontenc}
\usepackage{siunitx}
\usepackage{braket}
\usepackage{balance}

\newcommand{\DCS}{\left[  \frac{ {\rm d}\sigma} {{\rm d} \omega}  \right]  ^{\beta}_{\alpha}}
\newcommand{\mr}{{\widetilde m}_1}

\begin{document}

\title{Stereodynamic control  of cold rotationally inelastic CO + HD collisions}

\author{Pablo G. Jambrina}
 \email{pjambrina@usal.es}
\affiliation{%
 Departamento de Qu\'{\i}mica F\'{\i}sica ,
Universidad de Salamanca, Salamanca, Spain\
}%

\author{James F. E. Croft}
 \email{j.croft@otago.ac.nz}
\affiliation{%
 Department of Physics, University of Otago, Dunedin
9054, New Zealand\
}%
\altaffiliation{%
 Dodd-Walls Centre for Photonic and Quantum
Technologies, Dunedin 9054, New Zealand\
}%

\author{Naduvalath Balakrishnan}
 \email{naduvala@unlv.nevada.edu}
 \affiliation{
 Department of Chemistry and
Biochemistry, University of Nevada, Las Vegas, NV 89154, USA \
}%

\author{F. Javier Aoiz}
 \email{aoiz@ucm.es}
 \affiliation{
 Departamento de Qu\'{\i}mica
F\'{\i}sica , Facultad de Ciencias Qu\'{\i}micas, Universidad Complutense de
Madrid , 28040 Madrid  , Spain
}%

\begin{abstract}
  Quantum control of molecular collision dynamics is an exciting emerging area of cold  collisions. Co-expansion of collision partners in a supersonic molecular beam combined
with precise control of their quantum states  and alignment/orientation using  Stark-induced Adiabatic  Raman  Passage allows exquisite stereodynamic control of the collision
outcome. This approach has recently been demonstrated for rotational quenching of HD in collisions with H$_2$, D$_2$, and He and D$_2$ by He. Here we illustrate this
approach for HD($v=0,\,j=2$)+CO($v=0,\,j=0$)$\to$HD($v'=0,\,j'$)+CO($v'=0,\,j'$) collisions through full-dimensional quantum scattering calculations at collision energies near
\SI{1}{\kelvin}. It is shown that the collision dynamics at energies between 0.01--1\,K  are controlled by an interplay of $L=1$ and $L=2$ partial wave resonances depending on
the final rotational levels of the two molecules. Polarized cross sections resolved into magnetic sub-levels of the initial and final  rotational quantum numbers of the two
molecules also reveal significant stereodynamic effect in the cold energy regime. Overall, the stereodynamic effect is controlled by both geometric and dynamical factors, with
parity conservation playing an important role in modulating these contributions depending on the particular final state.
\end{abstract}

\maketitle


\section{Introduction}

The extraordinary progress achieved in the last couple of decades in creating dense samples of cold
and ultracold molecules has transformed our ability to control and interrogate the outcome of
molecular
collisions~\cite{Rev_Carr08,Roman_book,Rev_Bala16,Rev_Bohn17,Toscano2020,KRb_2019_Science,KRb_2020_NatPhys}.
This progress has led to new applications of cold and ultracold molecules in precision molecular
spectroscopy, quantum sensing, quantum information and computation, and quantum control of chemical
reaction
dynamics~\cite{segev.pitzer.ea:collisions,hu.liu.ea:nuclear,son.park.ea:collisional,kendrick.li.ea:non-adiabatic}.
Ultracold molecules in their absolute rovibrational and motional ground states
 trapped in optical tweezers  allow the realization of quantum
engineering of molecular assembly for many-body
dynamics~\cite{weyland.szigeti.ea:pair,cairncross.zhang.ea:assembly}, new quantum matter with
exotic properties and molecular qubits for quantum computation and
simulation~\cite{Rev_Bohn17,Toscano2020,albert.covey.ea:robust}.

Ultracold molecules offer unique opportunities to explore molecular collisions in the deep quantum
regime. One  such elementary molecular processes is a rotation-translation energy
exchange in which a rotationally excited molecule undergoes relaxation (quenching) in collisions
with an atom and the energy released is transferred to the relative translation of the collision
partners. Such processes have been extensively studied in the literature for many neutral and ionic
molecular systems, including the simplest molecule
H$_2$~\cite{wan.yang.ea:collisional,wan.balakrishnan.ea:rotational}. Collisions of H$_2$ and HD
with He are important for modeling gas densities in astrophysical environments where
non-equilibrium populations prevail~\cite{RevModPhys.85.1021,wan.balakrishnan.ea:rotational}. At
thermal energies, many partial waves contribute and the collision outcome is generally less
sensitive to fine details of the interaction potential. However, at the lowest temperatures of
interest in the interstellar medium ($\sim$4~K), only a few partial waves contribute for light
systems such as He+H$_2$, He+HD, and H$_2$+H$_2$. In this regime, collision outcomes are severely
influenced by small changes in the interaction potential and isolated resonances due to tunneling
through angular momentum barriers.

Currently there is much interest in studying inelastic and reactive molecular collisions near
\SI{1}{\kelvin} as well as in the \si{\milli\kelvin} (cold) and \si{\micro\kelvin} (ultracold)
regimes~\cite{2017_Science_Perreault,2018_NatChem_Perreault,SARP_HD-He,err_SARP_HD-He,SARP_HeD2,2018_PRL_Croft,2019_JCP_Croft,2019_PRL_Jambrina,Lara_JCP15,Lara_PRA15,jambrina_PCCP20,HeHD_Morita_ultracold,tscherbul.kos:magnetic,devolder.tscherbul.ea:coherent,devolder.brumer.ea:complete}.
Quantum effects are  amplified in these regimes and quantum control of molecular collisions using
external electric and magnetic fields becomes
feasible~\cite{Rev_Carr08,Roman_book,Rev_Bala16,Rev_Bohn17,Toscano2020,KRb_2019_Science,KRb_2020_NatPhys}.
While such control is most promising in the ultracold regime where only a single partial wave
contributes, collision energies near \SI{1}{\kelvin} are also of interest as collision
outcomes are dominated by a few partial waves.
The energy regime between \SI{1}{\milli\kelvin}--\SI{1}{\kelvin} has been the focus of many experiments involving
co-expansion~\cite{Amarasinghe2017,Amarasinghe2020} and merged beam
techniques~\cite{Henson2012,Shagam2015,Klein2017} in which sensitive measurements of isolated
resonances have been reported for Penning ionization of molecules such as H$_2$ and HD  by rare gas
atoms in excited electronic states. The regime near \SI{1}{\kelvin} has also been the focus of a
series of experiments by Perreault et al.\ in which rotational quenching of HD by H$_2$, D$_2$ and
He has been
reported~\cite{2017_Science_Perreault,2018_NatChem_Perreault,SARP_HD-He,err_SARP_HD-He,SARP_HeD2}.
The experiment involves co-expansion of the molecular species in a supersonic beam combined with
selection of the initial orientation of the molecular rotational angular momentum through
Stark-induced adiabatic Raman Passage (SARP). The SARP method allows stereodynamic control of the
collision process by selecting a given projection ($m_j$) of the molecular rotational angular
momentum $j$ on the relative collision velocity vector or preparing a molecular state in a coherent
superposition of $m_j$ states. For collision partners such as H$_2$ and HD or HD/D$_2$ and He, the
co-expansion can achieve a narrow distribution of relative molecular velocities corresponding to
collision energies in the vicinity of 1 Kelvin, drastically limiting the number of angular momentum
partial waves. Yet, experimental results do not provide explicit energy resolution and  theoretical
studies are needed to identify specific partial wave resonances that contribute to distinct
features in the experimental angular distribution or collision
mechanism~\cite{Jambrina_2019,Jambrina_CS16}. Computational studies have been  critical in yielding
mechanistic insights into recent experiments on HD($v=1,\,j=2 \to v'=0,\,j'=0$) quenching by H$_2$
and He~\cite{2018_PRL_Croft,2019_JCP_Croft,2019_PRL_Jambrina,HeHD_Morita_ultracold}.

Calculations have also demonstrated that stereodynamic control extends to cases where there are
overlapping resonances from multiple partial waves~\cite{Morita_HCl-H2,jambrina_PCCP20} making
it possible to disentangle the resonance pattern. Moreover, calculations have also identified strong
stereodynamic preference in the $m_j-m_{j'}$ resolved integral and differential cross sections in
the ultracold s-wave regime for rotational quenching of HD by He~\cite{HeHD_Morita_ultracold}.

So far, the experiments on state prepared HD with He and H$_2$/D$_2$ involved no change in rotational levels of the collision partners (H$_2$/D$_2$) limiting the number of
partial waves in the outgoing channel and complexity of the collision dynamics. However, it is not clear whether stereodynamic control of the collision outcome is still possible
when both collision partners change their rotational states. Here, we consider HD($v=0,\,j=2$)+CO($v=0,\,j=0$)$\to$HD($v'=0,\,j'$)+CO($v'=0,\,j'$) collisions in which rotational
levels of both molecules are altered during the collision leading to more intricate collision dynamics. Moreover, and unlike HD+H$_2$/He systems, the interaction potential for
CO+H$_2$ is deeper and  more anisotropic, offering a more stringent case for stereodynamical control in the cold regime.

Collisions of molecular hydrogen with CO are important processes in astrophysical environments and
has attracted considerable experimental and theoretical attention in recent
years~\cite{Jankowski_1998,Potapov_2009,Yang_2010,Jankowski_Science_2012,Jankowsk_JCP_2012,Chefdeville_2012,Yang2015,Chefdeville_2015,FORREY2015,Costes_2016,Faure_2016}.
Its importance stems from the fact that CO is the second most abundant molecule in the interstellar
medium after H$_2$ and is often used as a tracer of H$_2$ in dense interstellar clouds due to its
dipole moment. Several theoretical studies have reported temperature dependent rate coefficients
for rotational and rovibrational transitions in CO due to H$_2$ collisions of interest in
astrophysical
media~\cite{Yang_2010,Chefdeville_2012,Yang2015,Chefdeville_2015,FORREY2015,Costes_2016,Faure_2016}.
Highly accurate measurements of CO rotational excitation cross sections by H$_2$ have also been
reported  allowing direct comparisons with theoretical
predictions~\cite{Chefdeville_2012,Yang2015,Chefdeville_2015,Costes_2016,Faure_2016}. The most
recent full-dimensional potential energy surface (PES) for the H$_2$-CO
complex~\cite{Jankowsk_JCP_2012,Faure_2016} have yielded rotational excitation cross sections in
close agreement with experiment~\cite{Chefdeville_2015,Costes_2016} as well as high accuracy
spectroscopic data~\cite{Potapov_2009,Jankowski_Science_2012}. While CO+H$_2$ collisions have been
extensively studied, CO+HD collisions have received limited attention, and we are not aware of any
prior theoretical studies. The experimental measurements have reported anomalously large rate
coefficients for vibrational relaxation of CO($v=1$) by HD compared to H$_2$ and D$_2$ collision
partners~\cite{ANDREWS1976,Drozdoski1976,TURNIDGE1994}. Here we focus on rotational relaxation of
HD($v=0,\,j=2$) by CO($v=0,\,j=0)$ in which the HD molecule is prepared in various stereodynamic
alignment/orientations.

\section{Theoretical Approach}

\subsection{Scattering calculations}
Scattering calculations were performed in full-dimensionality using a modified version of the TwoBC
code~\cite{krems}, which implements a time-independent close-coupling formalism yielding the
scattering $S$ matrix~\cite{arthurs.dalgarno:theory}. This approach has has been outlined in detail
elsewhere~\cite{quemener.balakrishnan.ea:vibrational,
quemener.balakrishnan:quantum,santos.balakrishnan.ea:quantum}. While excited vibrational levels are
included in the basis set we only examine transitions between rotations levels in the ground
vibrational manifold and as such, for convenience, we label each asymptotic channel by the combined
molecular state $\alpha=j_1 j_2$, where $j$ is the rotational quantum number. In this work the
subscript 1 refers to HD and 2 to CO. The integral cross section for state-to-state rotationally
inelastic scattering is given by,
\begin{eqnarray}\label{eqn:ics}
  \sigma_{\alpha \to \alpha'} &=& \frac{\pi}{(2j_1+1)(2j_2+1)k_\alpha^2}\,\, \sum_{J}\sum_{j_{12},j'_{12}} \sum_{L,L'} \\ && \nonumber (2J+1)|T^{J}_{\alpha
  Lj_{12},\alpha'L'j'_{12}}|^{2}\,,
\end{eqnarray}
where $k_{\alpha}^2=2 \mu E_{\rm coll}/\hbar^2$ is the square of the initial relative wave vector, $E_{\rm coll}$ is the collision energy, $\mu$ is the reduced mass, $T^J =
1-S^J$, $\vec{j}_{12} = \vec{j}_1 + \vec{j}_2$, $L$ is the orbital angular
momentum quantum number, and $J$ the total angular momentum quantum number where $\vec{J} = \vec{L}
+ \vec{j}_{12}$.

For the PES we used the recent high-accuracy 6D CO+H$_2$ potential reported by Faure et al.~\cite{faure.jankowski.ea:on, garberoglio.jankowski.ea:all-dimensional}. This
potential was chosen as it reproduces the proper physical inverse-power dependence with the intermolecular distance, $R$, at long range which is crucial for low energy
collisions. To account
for the difference in centre of mass between H$_2$ and HD a coordinate rotation was implemented as
described in~\cite{balakrishnan.croft.ea:rotational}. Jacobi vectors were employed to describe the
relative positions of the atoms with $\vec{r_1}(r_1, \hat{r}_1)$ denoting the vector connecting H
with D, $\vec{r_2}(r_2, \hat{r}_2)$ denoting the vector connecting C with O, and $\vec{R}(R,
\hat{R})$ denoting the vector joining the centers of mass of the two molecules. The angular
dependence of the potential was expanded as
\begin{equation}
  U(\vec{r_1}, \vec{r_2}, \vec{R}) = \sum_\lambda A_\lambda(r_1, r_2, R)Y_\lambda(\hat{r}_1, \hat{r}_2, \hat{R}),
\end{equation}
with
\begin{eqnarray} \nonumber
  Y_\lambda(\hat{r}_1, \hat{r}_2, \hat{R}) &=& \sum_m \braket{\lambda_1 m_1 \lambda_2 m_2| \lambda_{12} m_{12}} Y_{\lambda_1
  m_1}(\hat{r}_1) \\ && \times Y_{\lambda_2 m_2}(\hat{r}_2)Y^*_{\lambda_{12} m_{12}}(\hat{R}),
\end{eqnarray}
where $\lambda \equiv \lambda_1\lambda_2\lambda_{12}$ and $m \equiv m_1 m_2 m_{12}$. For the
scattering calculations $\lambda_1$ was restricted to 0--4 while $\lambda_2$ was restricted to
0--8.

The coupled channel equations were propagated from 2 to \SI{92}{\bohr} with a radial step size of
\SI{1.25e-1}{\bohr} using a log-derivative method~\cite{manolopoulos:improved}. The number of
points in the radial coordinate for each dimer for the discrete variable representation was 18; the
number of points in the angular coordinate $\theta$ between $\vec{R}$ and $\vec{r}$ for each dimer
for the Chebyshev quadrature was 12; the number of points in the dihedral angle between
$\theta_1$ and $\theta_2$ for the Gauss-Hermite quadrature was 8. The quadratures are the same as
used by Yang et al.\ to study H$_2$ + CO collisions~\cite{Yang2015}. The basis set for the CO dimer
included vibrational levels 0 and 1 with rotational levels up to 8 and 2 respectively, while for HD
rotational levels up to 4 were included. Scattering calculations were performed for each parity for
$J \leq$ 12.

\begin{figure}[h]
\centering
  \includegraphics[height=6.5cm]{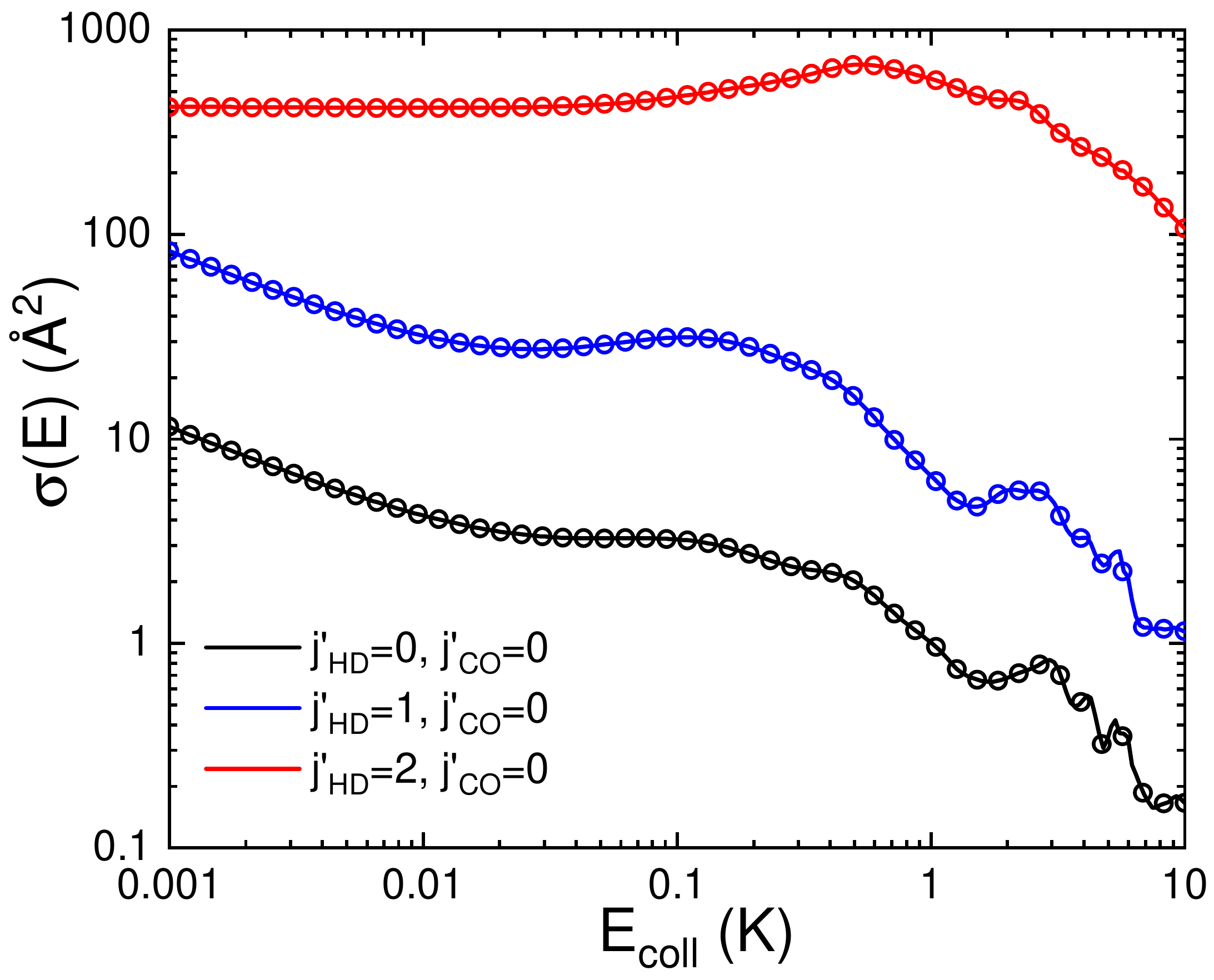}
  \caption{Comparison of the production basis set with an expanded basis
  to check convergence.
  Integral cross sections for the  HD($j_{\rm HD}$=2) + CO($j_{\rm CO}$=0) $\rightarrow$ HD($j'_{\rm HD}$=0-2) + CO($j'_{\rm CO}$=0)  transitions are shown,
  the solid lines correspond to the production basis set while the
  circles correspond to the expanded basis.
  It can be seen that no difference can be observed at the scale shown and the
  low energy behaviour is correctly reproduced.
  }\label{fig:convergence}
\end{figure}

To check convergence with respect to the basis set and radial propagation the integral cross
section was computed for the 20 $\to$ 00, 10, and 20 transitions with $J \leq 5$  using an expanded
basis: $\lambda_1$ was restricted to 0--10 while $\lambda_2$ was restricted to 0--6; The coupled
channel equations were propagated from 1.5 to \SI{102}{\bohr} with a radial step size of
\SI{1.05e-1}{\bohr}; The basis set for the CO dimer included vibrational levels 0 and 1 with
rotational levels up to 10 and 4 respectively, while for HD only the ground vibrational level was
included with rotational levels up to 6. Fig.~\ref{fig:convergence} compares the integral
cross-sections computed using both basis sets. The solid lines correspond to the ``production''
basis set described in the previous paragraph while the black crosses on a course grid correspond
to the expanded basis, it can be seen that no difference can be observed at the scale shown and the
low energy behaviour is correctly reproduced.
\subsection{Stereodynamics}

To compute the differential cross sections (DCS) we need first to compute the scattering amplitude ($f_{m_1, m_2 \rightarrow m'_1, m'_2
}$). From a given $\alpha \rightarrow \alpha'$ transition, the scattering amplitude in the helicity representation is given by~\cite{schaefer_79}:
\begin{align}
&f_{m_1, m_2 \rightarrow m'_1, m'_2 } = \frac{1}{2 k_{\alpha}} \sum_J (2J+1) \nonumber \\
&\sum_{j_{12},j'_{12},L,L'} i^{L - L' +1 } \,  d^J_{m_{12},m'{12}} (\theta) T^J_{\alpha L j_{12} \alpha' L' j'_{12}} \nonumber \\
& \times
\langle j'_{12} m'_{12}, J -m'_{12} | L' 0  \rangle \langle j_{12} m_{12}, J -m_{12} | L 0  \rangle \nonumber \\
&\times  \langle j'_{1} m'_{1},  j'_{2} m'_{2}  | j'_{12} m'_{12} \rangle
\langle j_{1} m_{1},  j_{2} m_{2}  | j_{12} m_{12} \rangle  \label{f}
\end{align}
where $m_{12} = m_1+m_2$, $m'_{12} = m'_1+m'_2$ (otherwise the last two Clebsch Gordan coefficients
are zero), and where $\alpha$ and $\alpha'$ have been omitted for the  sake  of clarity. In Eq.~\ref{f},
$\theta$ is the scattering angle,  $d^J_{m_{12},m'_{12}}(\theta)$ is Wigner's reduced rotation matrix, and
$\langle \ldots | \ldots\rangle$ is a Clebsch-Gordan coefficient. The DCS can be calculated as:
\begin{eqnarray} \nonumber
  \mathrm{DCS} &\equiv& \frac{d\sigma}{d\omega} = \frac{1}{(2 j_1 + 1) (2 j_2 + 1)} \\ &\times & \sum_{m_1,m_2}
  \sum_{m'_1,m'_2} f^*_{m_1, m_2 \rightarrow m'_1, m'_2 } f_{m_1, m_2 \rightarrow m'_1, m'_2 }
\end{eqnarray}
Similarly, three-vector correlations can be calculated in terms of the polarization dependent differential cross sections (PDDCS)~\cite{Aldegunde_JCPA05}.
Specifically, for the $\bm{k}$-${\bm j}_1$-${\bm k}'$ correlations,  where $\bm k$ and $\bm k'$ are the initial
and final relative velocities,  the corresponding reactant's PDDCS (or $j$-PDDCSs), $U^{(k)}_q(\theta)$, can be calculated as follows~\cite{Aldegunde_JCPA05}:
\begin{align} \label{ukq}
U^{(k)}_q (\theta)& = \frac{1}{(2 j_1 + 1) (2 j_2 + 1)}
 \\ \nonumber & ~~\times \sum_{m_1} \sum_{m_2} \sum_{m'_1,m'_2} f^*_{m_1, m_2 \rightarrow m'_1, m'_2 }
f_{m_1 + q, m_2 \rightarrow m'_1, m'_2 } \\ \nonumber & ~~\times \langle j_1 m_1, k q | j_1 m_1+q \rangle
\end{align}
where it is assumed  that reactant 1 (HD) is polarized, while reactant 2 (CO) is unpolarized. These PDDCSs are those that are needed to
simulate a SARP experiment where one of the reactants is polarized.

As described in Ref.~\citenum{Aldegunde_JCPA05}, if one of the reactant partners  is prepared in a directed state, $|j ~m=0\rangle$, its
internuclear axis can be aligned along the laboratory quantization axis, which is usually the light polarization vector~\cite{KL:JPC87}.  It is
possible then to change the direction of the quantization (laboratory-fixed) axis with respect  to the scattering frame, defined by $\bm k$
and $\bm k'$. The various directions of the polarization vector  lead to different relative geometries of the reactants. In particular, the
internuclear axis distribution for a given preparation is given by~\cite{Blum,Aldegunde_JCPA05}
\begin{align}\label{portrait}
P(\theta_r, \phi_r)= &\frac{1}{4\pi }\sum_{k} \sum_{q=-k}^{k} \,  (2k+1) \Big[ {\cal A}^{(k)}_0 C_{kq}(\beta, \alpha) \Big]  \nonumber \\
\times  C^{\, *}_{kq}(\theta_r, \phi_r) \,,
\end{align}
where $\beta$ and $\alpha$ are the polar and azimuthal angles that define the direction of the laboratory
quantization axis with respect to the scattering frame, $\theta_r$ and $\phi_r$ define the direction of the
relevant internuclear axis in the scattering frame.   $C_{kq}(\beta, \alpha)$ (and $C_{kq} (\theta_r,
\phi_r)$) are the modified spherical harmonics, and ${\cal A}^{(k)}_0$ are the polarization parameters that
define  the preparation in the laboratory frame.

In this scenario, the DCS for a given preparation of $\bm{j_1}$ is:
\begin{equation} \label{dcsalphabeta}
 \DCS  = \sum_{k=0}^{2 j_1} \sum_{q=-k}^{k} (2 k + 1) {\cal A}^{(k)}_0  \,
 U^{(k)}_q (\theta)  C^*_{kq} (\beta, \alpha).
\end{equation}

To obtain the integral cross section for the different experimentally achievable preparations, it is necessary to integrate $\displaystyle \DCS$ over both $\theta$ and the
azimuthal angle (for details see the appendix). Accordingly, the observable cross section depends only on $\beta$ and will be denoted as $\sigma^{\beta}$.

As shown in the appendix, if the initial rotational state for reactant 1  in the laboratory frame is the pure $|j\,m=0\rangle$ state,  hence  the
polarization parameters are given  ${\cal A}^{(k)}_0 = \langle j 0\,j 0 | k 0\rangle$ and it is possible to express the $(\beta, \alpha)$ DCS
directly in terms the scattering amplitudes:
 \begin{align}
 \DCS =& \frac{1}{2j_2+1}\, \sum_{m'_1 m'_2} \sum_{m_2}  \nonumber \\
 &\Big| \sum_{m_1} C_{j_1 m_1}(\beta, \alpha)  f_{m_1, m_2 \rightarrow m'_1, m'_2 }  \Big|^2
 \label{dcsalphabeta2}
\end{align}

\section{Results and Discussion}

\begin{figure}
  \centering
  \includegraphics[width=1.0\linewidth]{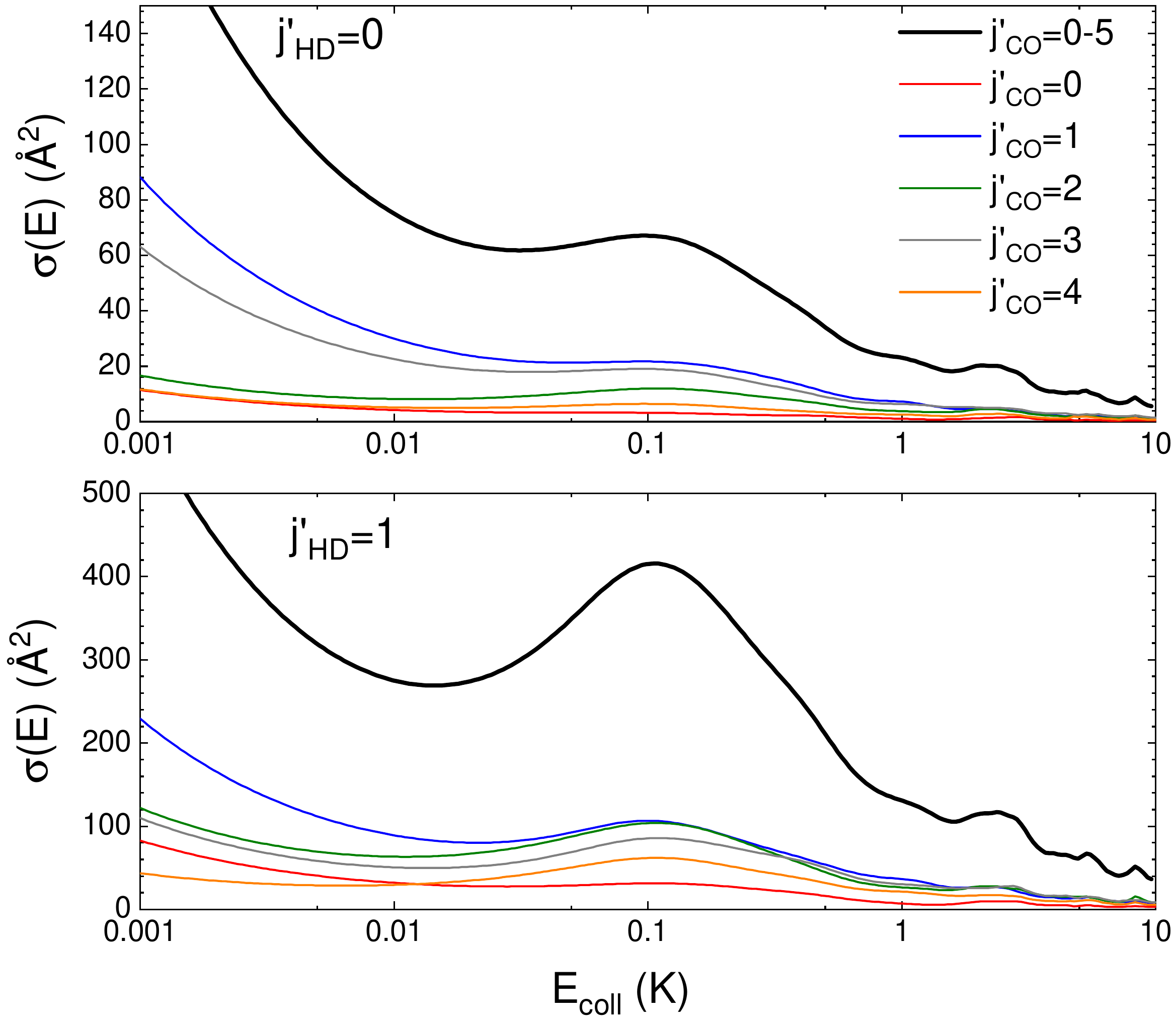}
  \caption{Cross section as a function of the collision energy for  HD($j_{\rm HD}$=2) + CO($j_{\rm CO}$=0)  $\rightarrow$ HD($j'_{\rm HD}$=0,1) + CO($j'_{\rm CO}$=0--5)
  collisions. Results for $j'_{\rm HD}$=0 are shown in the top panel and those for $j'_{\rm HD}$=1 are shown in the
  bottom panel.   }\label{Fig2}
\end{figure}

We will start this section showing the excitation function (cross section as a function of the collision energy) for the different final
rovibrational states that can be produced in the collisions between HD($v$=0,$j$=2) and CO($v$=0,$j$=0) at cold energies, between 1~mK
and 10~K. Due to the large difference in the rotational  constants of HD and CO ($B_{\rm e}$=64.2\,K and 2.8\, K for HD and CO, respectively) only $j'_{\rm HD}$=0-1 are
energetically accessible for HD but many different CO rotational states can be populated in this energy regime. Throughout this manuscript, we will divide
the final states according to the value of $j'_{\rm HD}$. The first endoergic channel corresponding to ($j'_{\rm HD}$=2, $j'_{\rm CO}$=1)
opens above 1 K, and has been omitted in our discussions.

\begin{figure*}
  \centering
  \includegraphics[width=1.0\linewidth]{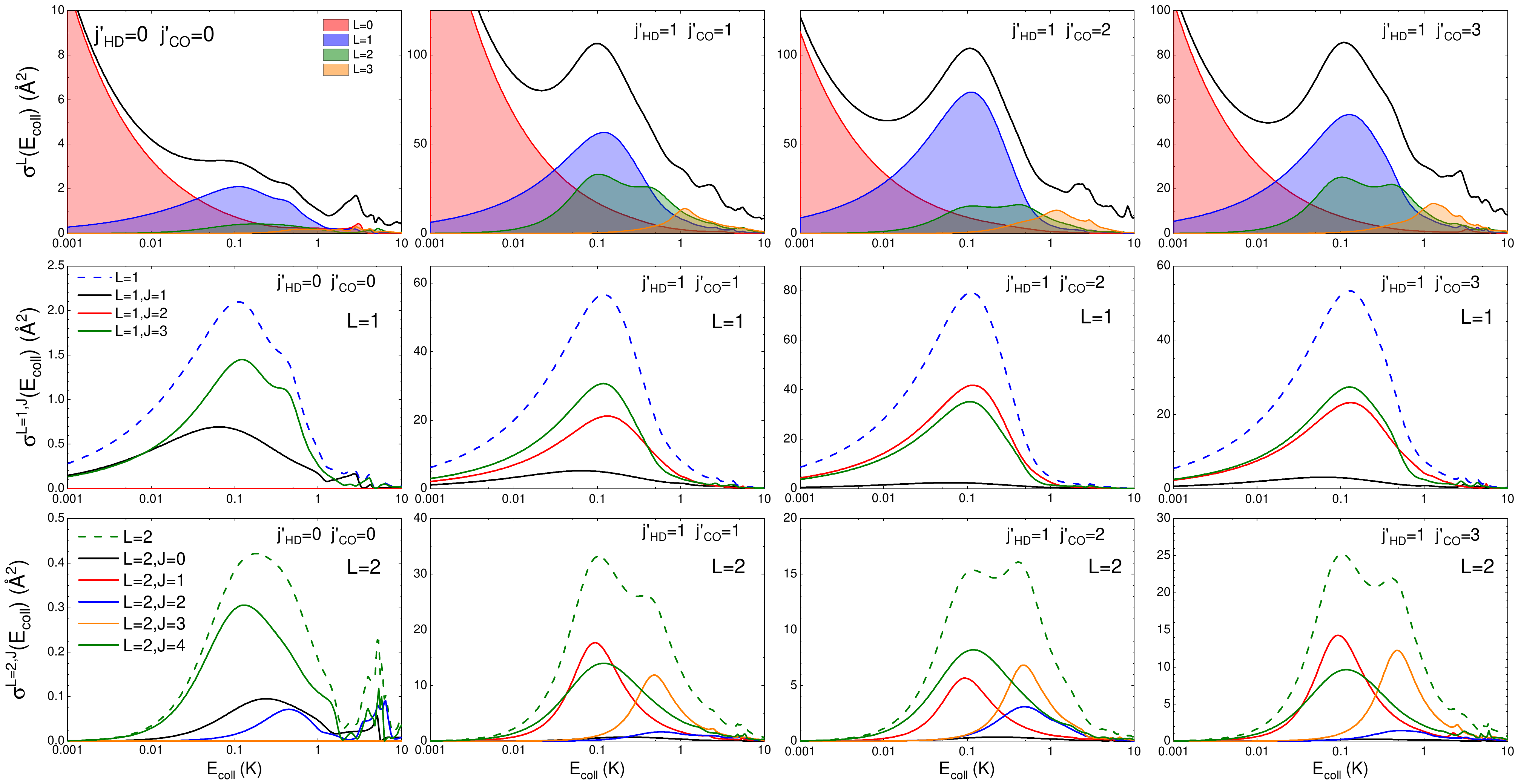}
  \caption{Total and partial cross sections as a function of the collision energy for the  HD($j_{\rm HD}$=2) + CO($j_{\rm CO}$=0)
  $\rightarrow$ HD($j'_{\rm HD}$) + CO($j'_{\rm CO}$) collision and four specific final states (see text for further details). The contribution of
  each $L$ value is shown in the top panel and  $\sigma(E_{\rm coll};J,L)$ are shown in the middle ($L$=1) and bottom ($L$=2) panels. For
  symmetry reasons  the number of possible ($J$, $L$)  combinations depend on the final state considered.}
  \label{Fig3}
\end{figure*}

Results displayed in Fig.~\ref{Fig2} show that collisions leading to $j'_{\rm HD}$=1 ($\Delta{j_{\rm HD}}$ =--1) have larger cross sections than those leading to $j'_{\rm HD}$=0
($\Delta{j_{\rm HD}}$ = --2). For a given  $j'_{\rm HD}$, collisions leading to  $j'_{\rm CO}$=1 are predominant, especially for $E_{\rm coll}<$ 10~mK, but that difference drops
as the energy rises. This indicates that HD rotational energy is not efficiently transferred to CO. The most relevant feature of the excitation function is the presence of a
resonance at around 0.1~K. The resonance peak is especially noteworthy for $j'_{\rm HD}$=1, but it is present for almost every final state.

To understand the origin of the resonance, and to determine the extent of control that can be achieved, we will focus on 4 different final
states: ($j'_{\rm HD}$=0, $j'_{\rm CO}$=0) for which the effect of the resonance, if any, is almost negligible, and ($j'_{\rm HD}$=1, $j'_{\rm
CO}$=1), ($j'_{\rm HD}$=1, $j'_{\rm CO}$=2), ($j'_{\rm HD}$=1, $j'_{\rm CO}$=3), the three states that are preferentially formed around the
energy of the resonance. In the top panels of  Fig. \ref{Fig3} we show the cross sections for these four states, and also the contribution of
$L$=0--3 to the cross section. As  expected, at low energies only $L$=0 contributes to the cross section, but with increasing collision
energies contributions from other $L$ become important. Around 0.1~K, the energy of the resonance peak, $L$=1 is the partial wave with the largest cross section, even for
($j'_{\rm HD}$=0, $j'_{\rm CO}$=0). Interestingly, the contribution of $L$=2 is rather different: for ($j'_{\rm HD}$=1, $j'_{\rm CO}$=1, 3) the  contribution from  $L$=2 to the
resonance is about 50\% that for $L$=1; however, for ($j'_{\rm HD}$=1, $j'_{\rm CO}$=2), ($j'_{\rm HD}$=0, $j'_{\rm CO}$=0)  its relevance is significantly smaller. We can
anticipate that the analogies between ($j'_{\rm HD}$=1, $j'_{\rm CO}$=1,3) will also manifest when we analyze both $\sigma^{J,L}$ and their behavior upon alignment of
$\bm{j}_{\rm HD}$.

In the two lower panels of  Fig.~\ref{Fig3}, we show  the partial cross sections for a given total and orbital angular momentum values. The results for  $L$=1 and $L$=2 are
displayed  in the middle and bottom panels, respectively. The sum over all $J$, i.e.\, $\sigma(E;L)$, are shown as dashed lines.  Parity conservation implies that not all ($L$,$J$)
combinations are possible for ($j'_{\rm HD}$=0, $j'_{\rm CO}$=0); as an instance,  ($L$=1, $J$=2) is forbidden. For this state and $L$=1,  $J$=1 and $J$=3 show similar cross
sections  at energies below the resonance,  but  $J$=3 prevails around the resonance.  For the other three final states, the contribution from ($L$=1, $J$=1) is very minor, while
those from ($L$=1, $J$=2) and ($L$=1, $J$=3) are similar. It is worth mentioning that while for ($j'_{\rm HD}$=1, $j'_{\rm CO}$=2),  $\sigma(E;J=2,L=1)$ is larger,  for $j'_{\rm
CO}$=1,3 $\sigma(E;J=2,L=3)$ prevails.

For $L$=2 differences between the four studied final states are substantial. For ($j'_{\rm HD}$=0,  $j'_{\rm CO}$=0) parity conservation ($P=(-1)^{j_{\rm HD}+j_{\rm CO}+L}$)
forbids  the ($J$=1,3) channels, and the dominant term is ($L$=2, $J$=4).  For $j'_{\rm HD}$=1 states, the relative cross section of ($L$=2, $J$=4) is similar, and these peaks
coexist with those observed for $J$=1,3. For $j'_{\rm HD}$=1, the peak associated with $\sigma^{L=2}$ exhibits  a double maximum, the first  associated with  $J$=1, and 4,
and the second associated with $J$=2, and 3. The position of these peaks does not depend on the final state,  although the relative contribution of $L$=2 to the resonance plays
an important role (see above). For the inelastic collisions between H + HF Jambrina et al.~\cite{jambrina_PCCP20}  also observed different maxima for a given $L$, each of them
associated with a different value of $J$.  However, in the present case,  the peak associated  with a given $J$ occurs at the same energy and  their position does not depend on
the final state.

 It is also worth emphasizing  that the relevant intensity of the partial cross sections for ($L$=2, $J$=1,3)  depends on the final state. For ($j'_{\rm HD}$=1, $j'_{\rm CO}$=1,3)
the  $J$=1 peak is higher than that of $J$=3, especially for  $j'_{\rm CO}$=1. On the contrary, for ($j'_{\rm HD}$=1,  $j'_{\rm CO}$=2) the  $J$=3 peak is more intense than
$J$=1   with a simultaneous increase in the intensity of the $J$=2 peak. As a result, the second maxima associated with $L$=2 is slightly higher than the first one, unlike what
was observed for $j'_{\rm CO}$=1,3. Again, the overall behavior of the $j'_{\rm CO}$=1,3 partial cross sections is different from that of  $j'_{\rm CO}$=2.

 \begin{figure}
  \centering
  \includegraphics[width=1.0\linewidth]{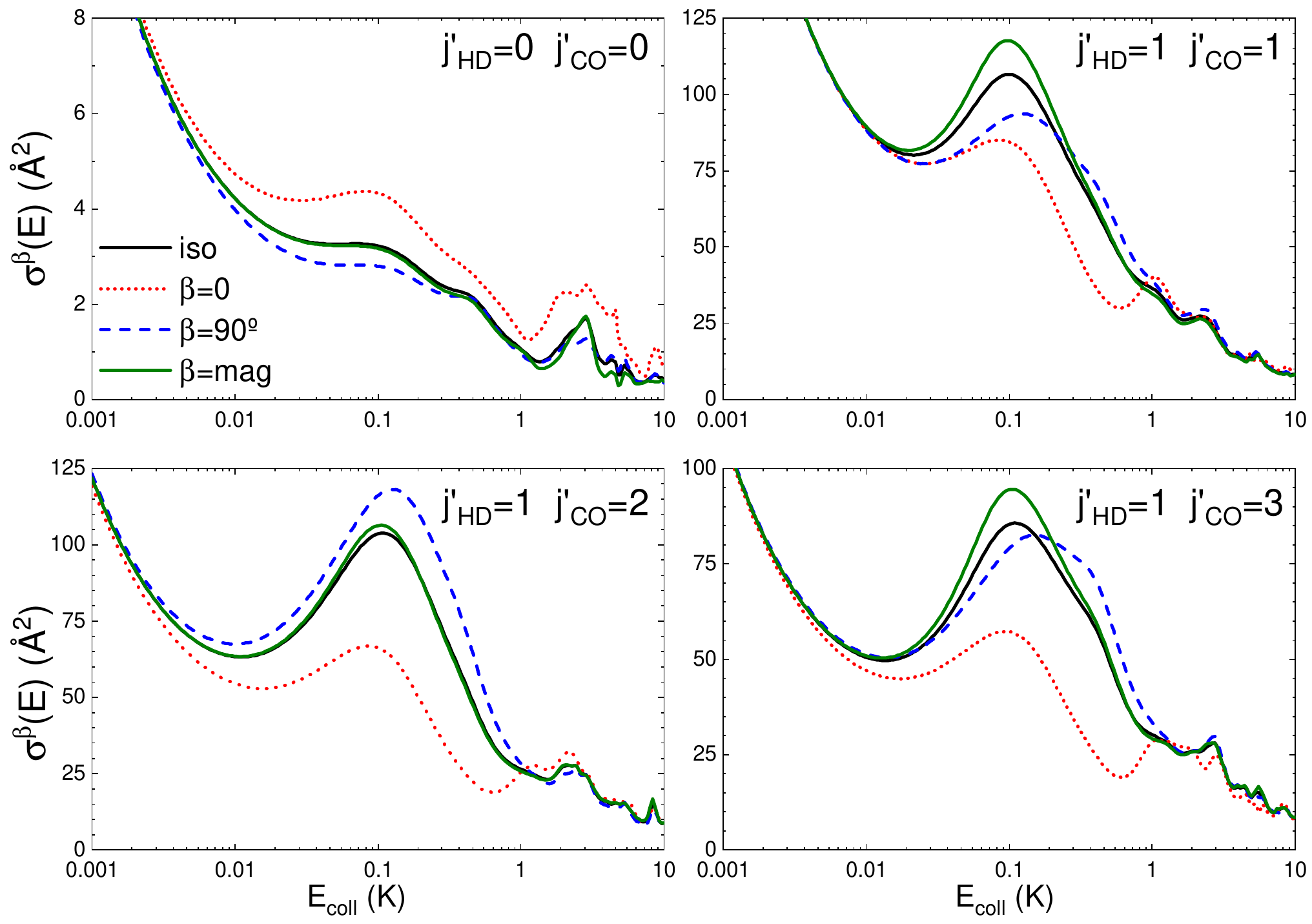}
  \caption{Cross section as a function of the collision energy for the  HD($j_{\rm HD}$=2)  + CO($j_{\rm CO}$=0)  $\rightarrow$ HD($j'_{\rm HD}$) + CO($j'_{\rm
  CO}$)   collisions for four specific final states and three different preparations of the HD intermolecular axis,
  $\beta=$0$^{\circ}$ (dotted red line),   $\beta=$90$^{\circ}$ (dashed blue line), and the magic angle (solid olive line). The isotropic
  preparation (in the absence of external alignment) is shown in black }
  \label{Fig4}
\end{figure}

To see if the intensity  of the resonance peaks can be tuned by selecting the direction of the HD rotational angular momentum (hence of the internuclear axis)  prior to the
collision, we calculated the excitation function for different values of $\beta$, the angle between the polarization vector of the SARP laser and the initial relative velocity vector
(see Fig.~\ref{Fig4}). With $\beta$=0$^{\circ}$ collisions will be preferentially head-on, while $\beta$=90$^{\circ}$ implies a side-on geometry. In fact, $\beta$=0$^{\circ}$ is
equivalent to selecting $m_1=m_{\rm HD}$=0. Between these two geometries, we will also calculate the excitation functions for $\beta={\rm mag}$ (the magic angle,
54.74$^{\circ}$, for which contribution of the $U^{(2)}_0)(\theta)$ term in Eq.~\eqref{dcsalphabeta} vanishes).

The alignment-dependent excitation functions display three different patterns depending on the final state considered, as shown in Fig.~\ref{Fig4}. For ($j'_{\rm HD}$=0,
$j'_{\rm CO}$=0),  for which the effect of the resonance is very minor,  the cross section is clearly enhanced for $\beta$=0$^{\circ}$ (head-on) collisions, in particular around
the 0.1~K resonance, as well as above 1~K. It should be emphasized that as $E_{\rm coll} \rightarrow$ 0, the integral  cross section  becomes insensitive to changes in the
relative alignment of the reactants.  \cite{AAMSA_JCP2006} In fact, if we only had $L$=0 (s-wave) encounters, there would not be any stereodynamical preference regardless of
the initial and final state considered. Side-on ($\beta$=90$^{\circ}$) collisions lead to slightly smaller cross sections, but the effect is not as noteworthy as for head-on
encounters. If $\beta={\rm mag}$ is selected, the cross section is nearly the same as if HD were not aligned (isotropic distribution).

  \begin{figure*}
  \centering
  \includegraphics[width=1.0\linewidth]{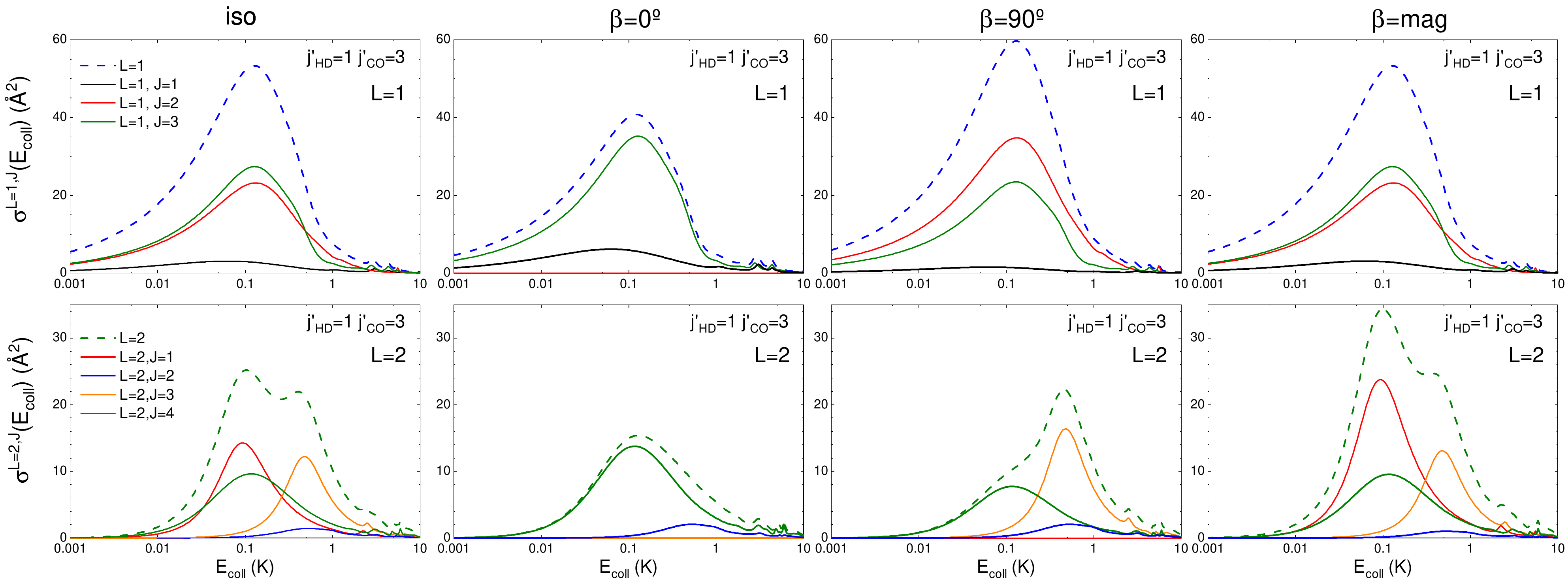}
  \caption{$\sigma(E;J,L)$ partial cross sections as a function of the collision energy for the HD($j_{\rm HD}$=2) + CO($j_{\rm CO}$=0)
  $\rightarrow$ HD($j'_{\rm HD}$=1) + CO($j'_{\rm CO}$=3)  collision for different preparations of the HD intermolecular axis. Results for
  $L$=1 are shown in the top panels and those for $L$=2 are shown in the bottom panels.  Results for the
isotropic distribution are identical to those depicted in Fig.~\ref{Fig3} and are only shown here for the sake of comparison.}
  \label{Fig5}
\end{figure*}

The situation is different for $j'_{\rm HD}$=1 states, for which the resonance has a salient effect. For these states,  the stereodynamical control is strongly influenced by the
resonance, and at energies above and below the resonance the relative alignment of HD has a
negligible effect on the cross sections. Moreover, regardless of the final state, the cross section around the resonance drops  for $\beta$=0$^{\circ}$
collisions, reaching its minimum value at around 0.6~K, beyond which  it rises to  the isotropic value. Again, there is
a clear difference between  $j'_{\rm CO}$=1,3 and $j'_{\rm CO}$=2. For ($j'_{\rm HD}$=1, $j'_{\rm CO}$=2), the cross section at the
resonance is enhanced by  the  $\beta$=90$^{\circ}$ preparation, while $\beta={\rm mag}$ has only a very minor effect. However, for  $j'_{\rm
CO}$=1,3, it is $\beta={\rm mag}$ that leads to a significant enhancement of the cross section. $\sigma^{\beta=90^{\circ}}$, in turn, is
somewhat smaller at the resonance peak and is shifted towards slightly higher collision energies.

To understand the origin of the different stereodynamical preferences, we show in Fig.~\ref{Fig5} the $\sigma^{J,L}$ for the different HD alignments discussed above. In this
case, we will  focus on one particular final state, ($j'_{\rm HD}$=1, $j'_{\rm CO}$=3). $\beta$=0$^{\circ}$  leads to a smaller $L$=1 cross section. This is due to the absence of
($L$=1,$J$=2) term which is not compatible with $\beta$=0$^{\circ}$ (due to parity conservation). For $L$=2,  $\beta$=0$^{\circ}$ leads to a modest increase of the cross
section associated with $J$=2 and $J$=4, but it makes  the terms associated with $J$ odd vanish (again imposed by conservation of the parity). Altogether, it explains the
decrease of the  reactivity associated with $\beta$=0$^{\circ}$ for $j'_{\rm HD}$=1 states. It also explains why $\beta$=0$^{\circ}$ leads to an increase of the partial cross
section for ($j'_{\rm HD}$=0, $j'_{\rm CO}$=0). Conservation of parity  requires  that for ($j'_{\rm HD}$=0, $j'_{\rm CO}$=0) the S-matrix element associated with ($L$=odd,
$J$=even) or ($L$=even, $J$=odd) must be  zero,  regardless of  the collision partner's polarization. These are the elements that are zero for  $\beta$=0$^{\circ}$ (since they do
not contain $m_1$=0), so the only effect of this preparation is to enhance  the contribution of the elements that are not zero by parity conservation, hence leading to an
increase of the cross section for ($j'_{\rm HD}$=0, $j'_{\rm CO}$=0).

Back to ($j'_{\rm HD}$=1, $j'_{\rm CO}$=3), $\beta$=90$^{\circ}$ enhances the influence of ($L$=1, $J$=2) but decreases that of
($L$=1, $J$=1). The effect on $L$=2 is more important, as  $\beta$=90$^{\circ}$ is not compatible with ($L$=2, $J$=1). Consequently,
 states for which the latter term was important show smaller cross sections for $\beta$=90$^{\circ}$. Besides, the enhancement of the
($L$=2, $J$=3) element displaces the resonance peak to slightly higher energies. The reason behind the disappearance of ($L$=2, $J$=1) for
$\beta$=90$^{\circ}$ can be found in Eq.~\eqref{f}.  For $j_1$=2, $L$=2, and $J$=1, the second Clebsch-Gordan in Eq.~\ref{f} is zero unless
$m_{12}$=1. And for $m_{12}$=1  and $\beta$=90$^{\circ}$ the cross-section is necessarily zero (see Eq.~\ref{sigbeta} in the appendix).

Finally, $\beta$=mag is compatible with all possible combinations of ($J$, $L$) and the term that is affected the most by this preparation
is ($L$=2, $J$=1), whose cross section is significantly larger. As a result, cross sections for those final states for which ($L$=2,
$J$=1) is important are enhanced by a $\beta={\rm mag}$  preparation.

 The  main theme that emerges from these discussions is as follows. An anisotropic preparation of the reactants leads to the modification of the
intensity of each $J$--$L$ combination. However, how the stereodynamic preparation changes these terms depends on geometric factors and not on the final
state, or any dynamical aspects. Nevertheless,   the relative weight of every $J$--$L$ contributions in the isotropic distribution has a dynamical origin. In
particular, dynamical quantum effects such as resonances are responsible of sudden and important changes in the modulus of particular
S-matrix elements.  In the present case, the effect of the resonance is very sensitive to the final state considered, leading to different
stereodynamical preferences.  In other words, by measuring the cross section for different experimental preparations, it could be possible to
disentangle the importance of the different $J$--$L$  partial waves.

 \begin{figure*}
  \centering
 \includegraphics[width=0.80\linewidth]{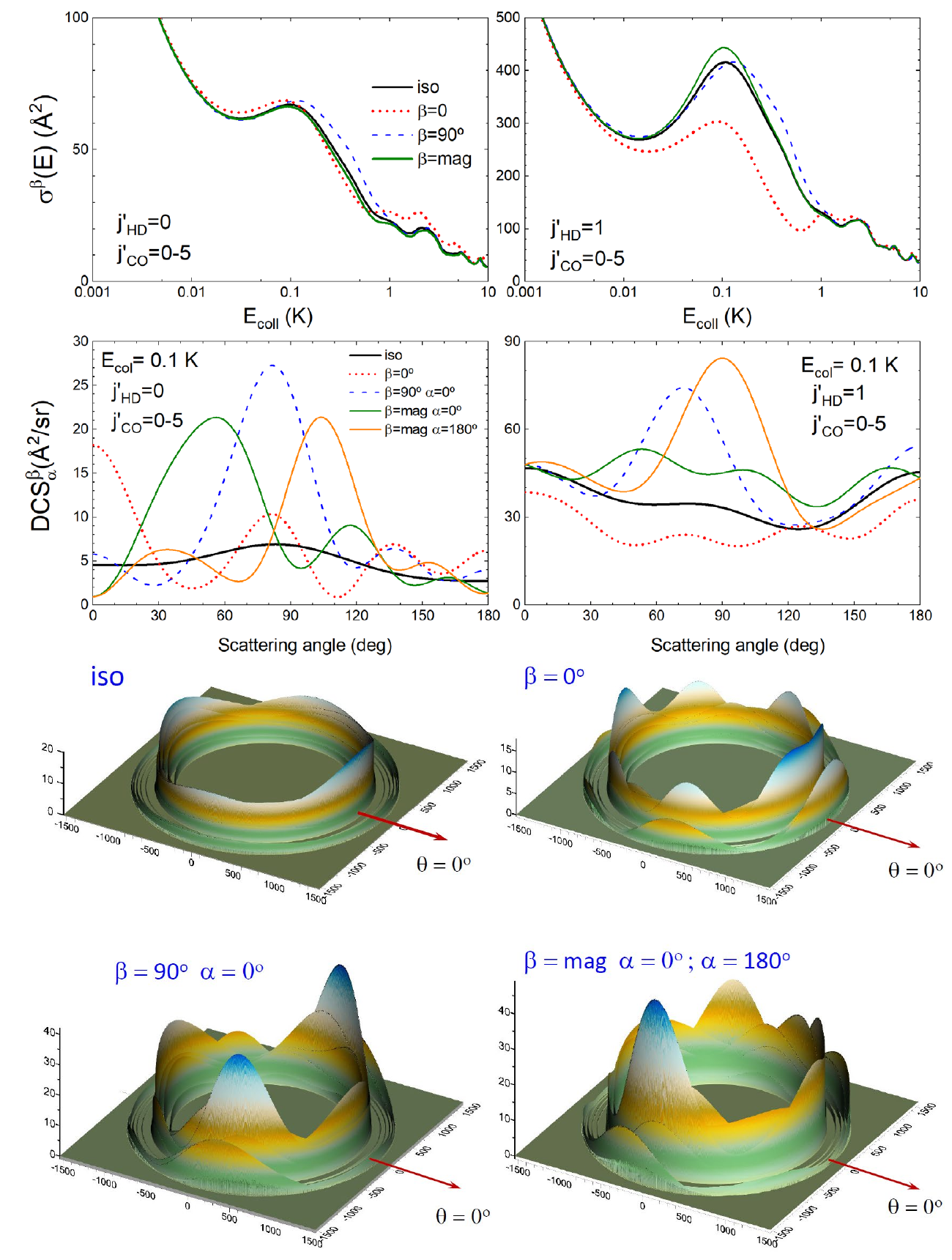}
 \caption{Top panels: Integral cross sections as a function of the collision energy for  HD($j_{\rm HD}$=2) + CO($j_{\rm CO}$=0)
 $\rightarrow$ HD($j'_{\rm HD}$=0,1) + CO($j'_{\rm CO}$)   collisions summed over all $j'_{\rm CO}$ and different stereodynamical preparations.
   Middle  panels: Differential cross sections at the peak energy of the resonance (0.1~K)  for $j'_{\rm HD}$=0 (left panel) and $j'_{\rm HD}$=1 (right panel)
summed over all  $j'_{\rm CO}$ states.  The abrupt changes in the intensity with the scattering angle for the various internuclear axis preparations are in stark contrast
with the relatively featureless shape of the isotropic DCS. Four Lower panels:  Scattering angle-recoil velocity polar maps at the same collision energy  for the
indicated internuclear axis preparations. Notice that DCS for $\beta=\mathrm{mag}$  and $\alpha$=0$^{\circ}$ differs from $\beta=\mathrm{mag}$  and
$\alpha$=180$^{\circ}$,  and therefore there is no azimuthal symmetry about
the incoming relative velocity. The preparation for $\beta$=90$^{\circ}$ has also no azimuthal symmetry but the DCS for $\alpha$=0$^{\circ}$ and $\alpha$=180$^{\circ}$
are the same.}\label{Fig6}
\end{figure*}

So far, we have only focused on how the different alignments affect the integral cross section. To see how they affect the DCS, in Fig.~\ref{Fig6} we show the DCS and the
scattering angle-recoil velocity polar maps at 0.1~K that could be experimentally measured detecting the HD rovibrational state.   As mentioned above, the energy difference
between two adjacent CO rotational states is considerably smaller than that between  consecutive HD rotational levels.  Therefore  in the polar maps we observe two rings: one
external (higher recoil velocities) associated with  $j'_{\rm HD}$=0 and an internal ring (lower recoil velocities) associated with $j'_{\rm HD}$=1. Since the cross section for
$j'_{\rm HD}$=1 is almost one order of magnitude larger than that for $j'_{\rm HD}$=0, the intensity of the internal ring is much higher. Along with the DCS and the polar
maps, in the two upper  panels of Fig.~\ref{Fig6} we show the alignment-dependent excitation function for $j'_{\rm HD}$=0, and 1 (summed over all final CO rotational states).
The alignment-dependent DCS are also shown in the two middle panels of~\ref{Fig6}.  As can be seen,   they exhibit a series of maxima that are not present in the  almost
featureless isotropic DCS, in particular for $j'_{\rm HD}$=0. For $j'_{\rm HD}$=1, the most salient features are the sideways peaks that can be observed for
$\beta$=90$^{\circ}$, $\alpha$=0$^{\circ}$  and  for $\beta$=mag, $\alpha$=180$^{\circ}$. These peaks can be also appreciated in the polar maps.   For
$\beta$=90$^{\circ}$, $\alpha$=0, 180$^{\circ}$ and $\beta$=mag, $\alpha$=0,180$^{\circ}$  there is a net increase of the cross section, for both $j'_{\rm HD}$.  As discussed
in prior studies~\cite{Liu:JCP14},  for $\beta$=mag, $\alpha$=0,180$^{\circ}$  the polar map is no longer symmetrical about the relative velocity, and the two  hemispheres are
different.  In the figure, a sideways peak is clearly appreciated for $j'_{\rm HD}$=1, in the ``$\alpha$=180$^{\circ}$ '' hemisphere, while a broader distribution is observed in the
``$\alpha$=0$^{\circ}$ '' hemisphere. It must be stressed that the integration over the scattering angle in the ($\beta, \alpha$)-DCSs does not correspond to  the
$\beta$-dependent cross sections shown in the two upper panels. Except for $\beta$=0, there is no azimuthal symmetry and, as shown in the Appendix, to reproduce the
$\sigma^{\beta}$ cross sections integration over all possible values of $\alpha$ (the azimuthal angle) is also required.  As expected, the effect of the stereodynamical
preparation on the DCSs is much more significant  than on the integral cross section.

\begin{figure*}
  \centering
  \includegraphics[width=1.0\linewidth]{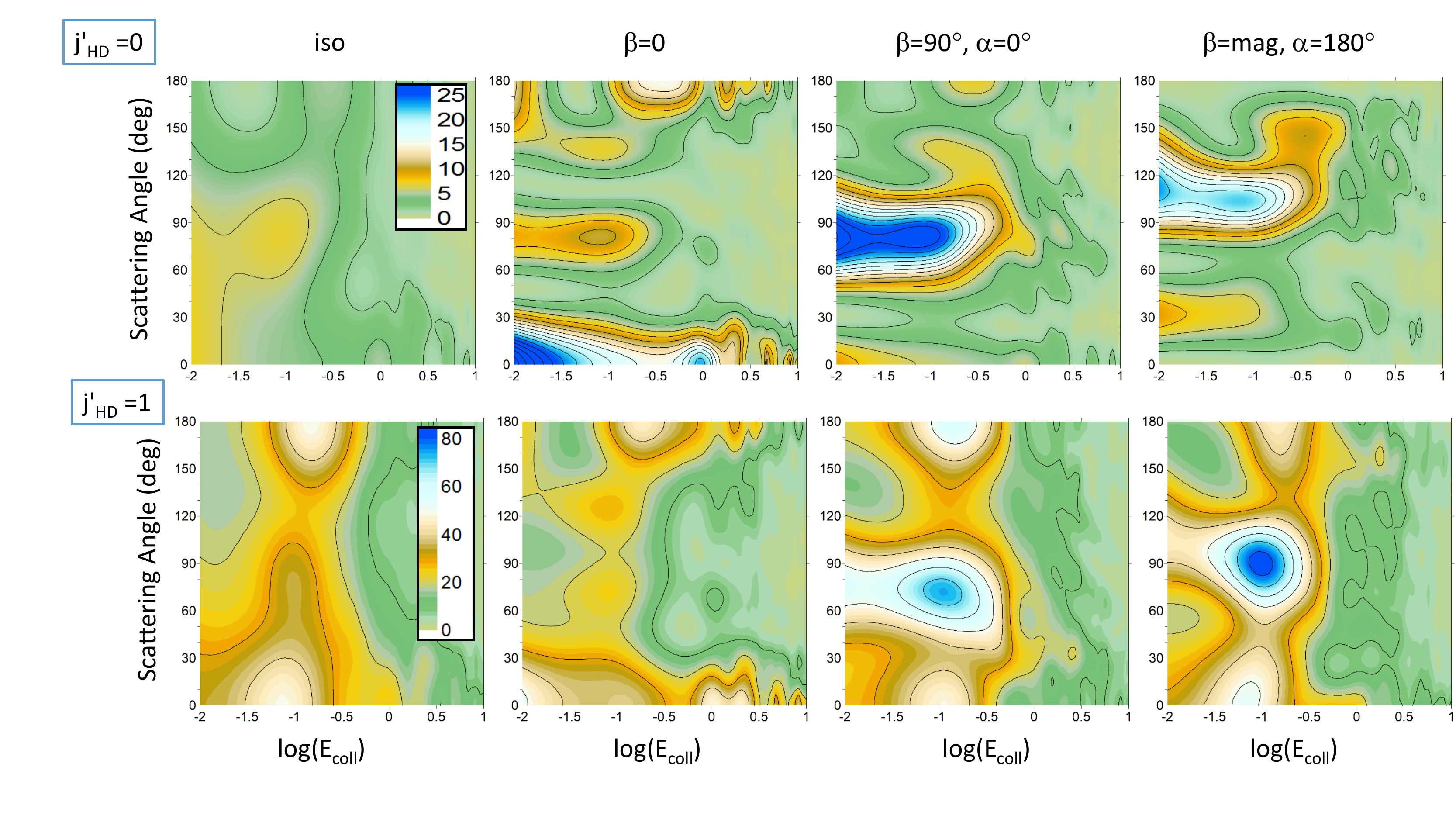}
  \caption{Contour plots showing the collision energy dependence of the DCS for the HD($j_{\rm HD}$=2) + CO($j_{\rm CO}$=0)
  $\rightarrow$ HD($j'_{\rm HD}$)+ CO($j'_{\rm CO}$)  collisions with different preparations of the HD internuclear axis. Results are summed
  over all $j'_{\rm CO}$ and are shown for $j'_{\rm HD}$=0 (top panels) and $j'_{\rm HD}$=1 (bottom panels).  }
  \label{Fig7}
\end{figure*}

To gain more insight on how specific features of the DCS is modified by the resonance, in Fig.~\ref{Fig7} we show the DCS calculated as a function
of the collision energy and the experimental preparation for both $j'_{\rm HD}$=0, and 1 (summing over all final CO rotational states). Even
though the ICS at low energies is largely independent of $\beta$, the features of the DCS changes significantly with $\beta$. Moreover, for $j'_{\rm
HD}$=1, we observe distinct features at the energies of the resonance, which are strongly influenced by changing the relative alignment of HD
angular momentum.

\section{Conclusions}
In this manuscript, we have studied the dynamics and stereodynamics of the inelastic collisions between HD($v$=0,$j$=2) and CO($v$=0,$j$=0) in the cold energy regime, i.e.\
for $E_{\rm coll}$ between 1~mK and 10~K. The main feature of the excitation functions (the energy dependence of the integral cross section) is the presence of a resonance at
$E_{\rm coll}$= 0.1~K, which is particularly relevant for $j'_{\rm HD}$=1 final states, that are more likely to be formed than their $j'_{\rm HD}$=0 counterparts.

Regardless of the final state considered, $L$=1 is the dominant partial-wave at the energy of the resonance peak. The relative population of $L$=2 at this energy depends on the
final state considered, and for most of the final states the resonance is observed for both partial waves. When the $L$--$J$ resolved cross sections ($\sigma^{L,J}$) are
examined, we observe that many $L$--$J$ combinations contribute to the resonance peak, and the relative intensity of these $L$--$J$ partial waves depends on the final state
considered.  It is the  interplay  between contributions from these $L$--$J$ partial waves that determines the preference towards one particular experimental preparation or
another. In particular, for ($j'_{\rm HD}$=1,  $j'_{\rm CO}$=2) higher cross section at the resonance are obtained for $\beta$=90$^{\circ}$ while $\beta$=mag is preferred for
($j'_{\rm HD}$=1, $j'_{\rm CO}$=1,3). In fact, the constraints imposed by the extrinsic alignment are similar to those imposed by parity conservation, making zero the cross
sections for some combinations of $L$ and $J$.

While at the integral cross section level, changing the polarization of $j_{\rm HD}$ only causes significant changes around the resonance, these changes are paramount when
the DCS are analyzed, and the DCS features a series of peaks, which are appreciable in a scattering angle-recoil velocity polar map and depend on the particular preparation
used.

As a whole, our results show that by tuning the polarization of one of the reactants it is possible to modify the effect of a resonance in the cold energy region for a process in
which the two partners change their rotational states. Moreover, since the $J$-$L$ partial waves are very dependent of the final state considered and will ultimately determine
the extent of stereodynamical control, it is possible to modify  to some extent the relative population of some of the product channels.  While the influence of the different
polarizations on the  $J$-$L$ partial wave is purely geometrical, the contribution  from these partial waves on the isotropic cross section is solely determined  by dynamics.
Therefore, the overall effect of the polarization on the intensity and width of the resonance will depend on both geometrical and dynamical effects.

\section*{Appendix}

The general expression for DCS for a given preparation of $\bm{j}_1$ while $\bm{j}_2$=0 or unpolarized,  is \cite{Aldegunde_JCPA05}
\begin{equation} \label{dcsalphabeta2}
 \DCS = \sum_{k=0}^{2 j_1} \sum_{q=-k}^{k} (2 k + 1) {\cal A}^{(k)}_0  \, U^{(k)}_q (\theta)  C^{\, *}_{kq} (\beta, \alpha).
 \end{equation}
Inserting  the expression for the PDDCS,  given by  Eq.~\eqref{ukq},  in Eq.~\eqref{dcsalphabeta2} yields,
\begin{eqnarray}\label{dcsalphabeta4}
 \DCS &=& \sum_k \sum_{q=-k}^{k} \Big(\frac{2 k + 1 }{2j_1+1} \Big) \,
 {\cal A}^{(k)}_0  C^{\, *}_{kq}(\beta,\alpha) \\ \nonumber & \times &
\sum_{m_1,\,\mr} Q_{m_1,\mr} \langle j_1 m_1 \,  k q| j_1\, \mr \rangle
\end{eqnarray}
where\cite{HWBJA:NC19,WHJAB:JPCA19}
\begin{equation}
 Q_{m_1\, {\widetilde m}_1}=  \frac{1}{2j_2+1}\,  \sum_{m'_1 m'_2}\,\, \sum_{m_2}
f_{m_1, m_2 \rightarrow m'_1, m'_2 }~ f^*_{\mr, m_2 \rightarrow m'_1, m'_2 }
\end{equation}
If the initial prepared state  in the laboratory frame is $|j_1\,\, 0\rangle$,  where it is assumed that $j$ is integer,  then ${\cal A}^{(k)}_0 = \langle j_1\, 0 \,  \,k \,0| j_1\, 0
\rangle$,
and
\begin{align}
\DCS = &\sum_{m_1,\mr} Q_{m_1,\mr} \, \sum_k \sum_{q=-k}^{k} \Big(\frac{2 k + 1 }{2j_1+1} \Big) \nonumber \\
 & \times \langle j_1\, 0 \,\,  k\, 0| j_1\, 0 \rangle \,  \langle j_1 \,m_1 \, \, k q| j_1 \, \mr \rangle \,C^{\,*}_{kq}(\beta,\alpha).
 \label{dcsalphabeta5}
\end{align}
Changing the order of the C.-G. coefficients~\cite{Zare}:
\begin{eqnarray}
\langle j_1 \, 0 \,\,  k\, 0| j_1\, 0 \rangle = (-1)^{j_1}  \Big(\frac{2 j_1 + 1 }{2k+1} \Big)^{1/2}  \langle j_1   0  j_1    0| k 0 \rangle
\end{eqnarray}
\begin{eqnarray}
\langle j_1 \, m_1 \,  \,k \, q| j_1 \, \mr \rangle &=& (-1)^{j_1-m_1}\, \Big(\frac{2 j_1 + 1 }{2k+1} \Big)^{1/2}\, \\ \nonumber & \times & \langle j_1\, m_1 \, \,  j_1\,  -\mr| k \, -q
\rangle
\end{eqnarray}
and taking into account that
\begin{align}
&\left[D^{k}_{q \, 0}(\alpha, \beta, 0)\right]^*=C_{kq}(\beta, \alpha),\\
&C^{\, *}_{kq}(\beta, \alpha)= (-1)^q C_{k -q}(\beta, \alpha)= (-1)^q \left[D^k_{-q \, 0}(\alpha, \beta, 0)\right]^*,
\end{align}
we obtain
\begin{align}\label{dcsalphabeta6}
\DCS =  & \sum_{m_1,\mr} (-1)^{2j_1-m_1+q} \,\, Q_{m_1,\mr} \, \sum_k
\langle j_1\, 0 \,\,  j_1 0| k \, 0 \rangle \, \nonumber \\
& \times  \langle j_1\,m_1 \, \, j_1 \, - \mr| k \, - q \rangle \, \left[D^k_{-q \, 0}(\alpha, \beta, 0)\right]^*
\end{align}
where $q=\mr - m_1$.

Considering the identity~\cite{Zare}:
\begin{eqnarray}\label{Dmatrix}
D^{J_1}_{M_1' M_1} \, D^{J_2}_{M_2' M_2} &=& \sum_{J_3} \langle J_1 M_1 \, J_2 M_2| J_3 M_3 \rangle \\ \nonumber & \times &
\langle J_1 M'_1 \, J_2 M'_2| J_3 M'_3 \rangle \,D^{J_3}_{M_3' M_3} \, ,
\end{eqnarray}
if $J_1=J_2=j_1$,  $J_3=k$,  $M_1=M_2=M_3=0$,  $M_1'=m_1$, $M_2'= - \mr$ and ${M_3'=m_1-\mr=-q}$
\begin{eqnarray}\label{Dmatrix2}
  \left[D^{j_1}_{m_1 0}\right]^* \, \left[D^{j_1}_{-\mr 0}\right]^* &=& \sum_{k} \langle j_1\, 0 \,\, j_1\, 0| k \, 0 \rangle \\ \nonumber & \times &
  \langle j_1 \, m_1 \, \, j_1 \,-\mr| k\,  - q \rangle \, \left[D^{k}_{- q \, 0}\right]^* ,
\end{eqnarray}
or
\begin{align}\label{Ckq12}
& (-1)^{\mr}\, C_{j_1 \, m_1}(\beta,\alpha) \, C^{\, *}_{j_1\, \mr}(\beta,\alpha)= \sum_{k} \langle j_1\, 0 \,\, j_1 \,0| k 0 \rangle \nonumber \\
&
\times \langle j_1 \,m_1 \,\,  j_1 \, - \mr| k \,  - q \rangle \, \left[D^{k}_{- q \, 0}\right]^* .
\end{align}
Substituting in Eqn.~\eqref{dcsalphabeta6} and bearing in mind that $(-1)^{2j_1-2m_1+2\mr} = +1$  for integer $j$, one obtains :
\begin{align}\label{dcsalphabeta7}
&\DCS  = \sum_{{m_1}, {\widetilde m}_1}  \,  Q_{m_1\, \mr} \,
C_{j_1 m_1}(\beta, \alpha)  C^*_{j_1 \mr} (\beta, \alpha) =  \nonumber \\
&  \frac{1}{2j_2+1}\, \sum_{m'_1 m'_2} \sum_{m_2}  \Big| \sum_{m_1} C_{j_1 m_1}(\beta, \alpha)  f_{m_1, m_2 \rightarrow m'_1, m'_2 }
\Big|^2.
\end{align}
\vspace{-0.5cm}

In those cases in which the experiment is carried out under conditions that imply  azimuthal symmetry,  Eq.~\eqref{dcsalphabeta7} ought to
be integrated over the azimuthal angle $\alpha$ if the $\bm k$-$\bm k'$ plane is taken as the reference or over $\phi-\alpha$  if a different
reference plane  is chosen.  The resulting expression is
\begin{align}
&\int_{0}^{2\pi} ~{\rm d}\alpha \DCS  = \nonumber \\
&\sum_{m_1, \mr}  2 \pi  \delta_{m_1, \mr} ~  Q_{m_1\, \mr} C_{j_1 m_1}(\beta, 0)  C_{j_1, \mr}
(\beta, 0) =\nonumber \\ & \qquad  2\pi \sum_{m_1} Q_{m_1, m_1} \Big[C_{j m_1} (\beta, 0) \Big]^2.
\label{intalpha}
\end{align}
Additional integration over $\cos\theta$ leads to  the `directional' integral cross section for one of reagents prepared with internuclear  axis
along $\beta$:
\begin{eqnarray}\label{sigbeta}
\sigma^{\beta} &=& \int_{-1}^{1} \, \int_{0}^{2\pi} ~{\rm d}(\cos \theta) \, {\rm d}\alpha \DCS \\ \nonumber &=& \sum_{m_1} \, \sigma_{m_1} \,\Big[C_{j_1 m_1}(\beta, 0)
\Big]^2\,,
\end{eqnarray}
where $\sigma_{m_1 }$ is the integral cross section for the $m_1$ state, averaged over $m_2$ and summed over $m_1'$ and $m_2'$
\begin{equation}\label{sigm}
\sigma_{m_1} = \frac{1}{2j_2+1} \, \sum_{m_1', m_2'} \sum_{m_2} \,\sigma_{m_1 m_2 \to  m_1' m_2'}.
\end{equation}

It  should be stressed  that Eqns.~\eqref{dcsalphabeta7}, and~\eqref{sigbeta} are only valid as long as the prepared state is $| j_1 \, 0\rangle$.

\section*{Conflicts of interest}
There are no conflicts to declare.

\section*{Acknowledgements}

PGJ and FJA thank Prof. Enrique Verdasco and Jesús Aldegunde for their support and help with the calculations. Funding by the Spanish Ministry of Science and Innovation
(grant   PGC2018-096444-B-I00) is  acknowledged. P.G.J. acknowledges funding by Fundaci\'on Salamanca City of Culture and Knowledge (programme for attracting scientific
talent to Salamanca). N.B. acknowledges support from NSF [Grant No. PHY-1806334] and ARO MURI [Grant No. W911NF-19-1-0283].

\balance


\providecommand*{\mcitethebibliography}{\thebibliography}
\csname @ifundefined\endcsname{endmcitethebibliography}
{\let\endmcitethebibliography\endthebibliography}{}

\end{document}